\documentstyle[preprint,epsfig,aps]{revtex} 
\tightenlines
\begin{document}
\draft
\preprint{\vbox{ \hbox{hep-ph/yymmdd}  
\hbox{FTUV/99-41} 
\hbox{IFIC/99-43} }}
\title{Status of the MSW Solutions of the Solar Neutrino Problem}
\author{M.\ C.\ Gonzalez-Garcia$^{1}$,
P. C. de Holanda$^{1,2}$,
C. Pe\~na-Garay$^{1}$ and
J.\ W.\ F.\ Valle$^{1}$
}
\address{\sl $^1$  Instituto de F\'{\i}sica Corpuscular -- C.S.I.C. \\
Universitat of Val\`encia,46100 Burjassot, Val\`encia, Spain\\
$^2$ Instituto de F\' {\i}sica Gleb Wataghin\\
Universidade Estadual de Campinas, UNICAMP
13083-970 -- Campinas, Brazil}
\maketitle
\begin{abstract}
\baselineskip 0.5 cm
We present an updated global analysis of two-flavor MSW solutions to
the solar neutrino problem.  We perform a fit to the full data set
corresponding to the 825-day Super--Kamiokande data sample as well as
to Chlorine, GALLEX and SAGE experiments. In our analysis we use all
measured total event rates as well as all Super--Kamiokande data on
the zenith angle dependence, energy spectrum and seasonal variation of
the events.  We compare the quality of the solutions of the solar
neutrino anomaly in terms of conversions of $\nu_e$ into active or
sterile neutrinos. For the case of conversions into active neutrinos
we find that, although the data on the total event rates favours the
Small Mixing Angle (SMA) solution, once the full data set is included
both SMA and Large Mixing Angle (LMA) solutions give an equally good
fit to the data.  We find that the best--fit points for the combined
analysis are $\Delta m^2=3.6 \times 10^{-5}$ eV$^2$ and $\sin^2
2\theta=0.79$ with $\chi^2_{min}=35.4/30$ d.~o.~f and $\Delta
m^2=5.1~\times 10^{-6}$ eV$^2$ and $\sin^2 2\theta=5.5 \times 10^{-3}$
with $\chi^2_{min}=37.4/30$ d.~o.~f.  In contrast with the earlier
504-day study of Bahcall-Krastev-Smirnov our results indicate that the
LMA solution is not only allowed, but slightly preferred. On the other
hand we show that seasonal effects, although small, may still reach 11\% 
in the lower part of the LMA region, without conflict with the
negative hints of a day-night variation (6\% is due to the
eccentricity of the Earth's orbit). In particular the best-fit LMA
solution predicts a seasonal effect of 8.5\%. For conversions into
sterile neutrinos only the SMA solution is possible with best--fit
point $\Delta m^2=5.0~\times 10^{-6}$ eV$^2$ and $\sin^2 2
\theta=3.\times 10^{-3}$ and $\chi^2_{min}=40.2/30$ d.~o.~f.  We
also consider departures of the Standard Solar Model (SSM) of Bahcall
and Pinsonneault 1998 (BP98) by allowing arbitrary $^8B$ and $hep$
fluxes. These modifications do not alter significantly the oscillation
parameters. The best fit is obtained for $^8B/^8B_{SSM}=0.61$ and
$hep/hep_{SSM}=12$ for the SMA solution both for conversions into
active or sterile neutrinos and $^8B/^8B_{SSM}=1.37$ and
$hep/hep_{SSM}=38$ for the LMA solution.
\end{abstract}
\newpage

\section{Introduction}

It is already three decades since the first detection of solar
neutrinos. It was realized from the very beginning that the observed
rate at the Homestake experiment \cite{homestake0} was far lower than
the theoretical expectation based on the standard solar model
\cite{SSMold} with the implicit assumption that neutrinos created in
the solar interior reach the earth unchanged, i.e. they are massless
and have only standard properties and interactions. In the first two
decades of solar neutrino research, the problem consisted only of the
discrepancy between theoretical expectations based upon solar model
calculations and the observations of the capture rate in the chlorine
solar neutrino experiment.  This discrepancy led to a change in the
original goal of using solar neutrinos to probe the properties of the
solar interior towards the study of the properties of the neutrino
itself.

From the experimental point of view much progress has been done in
recent years. We now have available the results of five experiments,
the original Chlorine experiment at Homestake \cite{homestake}, the
radio chemical Gallium experiments on pp neutrinos, GALLEX
\cite{gallex} and SAGE~\cite{sage}, and the water Cherenkov 
detectors Kamiokande~\cite{kamioka} and Super--Kamiokande
\cite{sk1,sk700}. The latter has been able not only to confirm the original 
detection of solar neutrinos at lower rates than predicted by standard
solar models, but also to demonstrate directly that the neutrinos come
from the sun by showing that recoil electrons are scattered in the
direction along the sun-earth axis. We now have good information on
the time dependence of the event rates during the day and night, as
well as a measurement of the recoil electron energy spectrum.  After
825 days of operation, Super--Kamiokande has also presented
preliminary results on the seasonal variation of the neutrino event
rates, an issue which will become important in discriminating the MSW
scenario from the possibility of neutrino oscillations in
vacuum~\cite{ourseasonal,v99}.

On the other hand there has been improvement on solar modelling and
nuclear cross sections. For example, helioseismological observations
have now established that diffusion is occurring and by now most solar
models incorporate the effects of helium and heavy element
diffusion~\cite{Bahcall:1997qw,Bahcall:1995bt}.  The quality of the
experiments themselves and the robustness of the theory make us
confident that in order to describe the data one must depart from the
Standard Model (SM) of particle physics interactions, by endowing
neutrinos with new properties. In theories beyond the Standard Model
of particle physics neutrinos may naturally have exotic properties
such as non-orthonormality~\cite{Schechter:1980gr}, flavour-changing
interactions~\cite{Hall:1986dx}, transition magnetic moments
~\cite{Schechter:1981hw} and neutrino decays~\cite{Schechter:1982cv},
the most generic is the existence of mass.  While many of these may
play a role in neutrino propagation and therefore in the explanation
of the data~\cite{allpossiblemechanisms} it is undeniable that the
most generic and popular explanation of the solar neutrino anomaly is
in terms of neutrino masses and mixing leading to neutrino
oscillations either in {\sl vacuum}~\cite{Glashow:1987jj} or via the
matter-enhanced {\sl MSW mechanism}~\cite{msw}.

In this paper we study the implications of the present data on solar
neutrinos in the framework of the two-flavor MSW solutions to the
solar neutrino problem.  We perform a global fit to the full data set
corresponding to 825 days of data of the Super--Kamiokande experiment
as well as to Chlorine, GALLEX and SAGE. In our analysis we use as
SSM the latest results from the most accurate calculation of neutrino
fluxes by Bahcall and Pinsonneault \cite{BP98} which incorporates the
new normalization for the low energy cross section
$S_{17}=19^{+4}_{-2}~{\rm eV~b}$ indicated by the recent studies at
the Institute of Nuclear Theory~\cite{Adelberger:1998qm}.  We have
also considered the possibility of departing from the SSM of BP98 by
allowing a free normalization of the $^8B$ flux and $hep$ neutrino
fluxes and we present the results we obtain when we treat these
normalization as free parameters.

We combined the measured total event rates at Chlorine, Gallium and
Super--Kamiokande experiments with the Super--Kamiokande data on the
zenith angle dependence, energy spectrum and seasonal variation of the
events. The goal of such analysis is not only to compare the quality of the 
solutions to the solar neutrino anomaly in terms of flavour oscillations of
$\nu_e$ into active or sterile neutrinos but also to study the weight
of different observables on the determination of the underlying
neutrino physics parameters as emphasized in Ref.~\cite{bks}.   
The outline of the paper is as follows.
In Sec.~\ref{sec:data} we present the basic elements that enter into
our calculation of the observables and the definitions used in the
statistical combination of the data.  Section
\ref{sec:fits} contains our results of the allowed (or excluded)
regions of oscillation parameters from the analyses of the different
observables. The results on the allowed regions from the combined
analysis of the total event rates is contained in
Sec.~\ref{fit:rates}.  The constraints arising from the
Super--Kamiokande searches for day--night variation of the event rates
are discussed in Sec.~\ref{fit:dn}.  In Sec.~\ref{fit:spec} we discuss
the information which can be extracted from the distortion of the
recoil electron energy spectrum measured by Super--Kamiokande. The
restrictions arising from the preliminary Super--Kamiokande data on
the seasonal variation of the event rates are studied in
Sec.~\ref{fit:sea}. In Sec.~\ref{fit:total} we present our results
from the global fit to the full data set and we determine the allowed
range of oscillation parameters which are consistent with all the
data. Finally in Sec.~\ref{conclu} we discuss the possible
implications of our results for future investigations.

Our results show that for oscillation into active neutrinos although
the data on the total event rates favours the SMA angle solution, once
the full data set is included both SMA and LMA give an equally good
fit to the data.  We find that the best--fit points for the combined
analysis are $\Delta m^2=3.6 \times 10^{-5}$ eV$^2$ and $\sin^2
2\theta=0.79$ with $\chi^2_{min}=35.4/30$ d.~o.~f. and $\Delta m^2=5.1
\times 10^{-6}$ eV$^2$ and $\sin^2 2\theta=5.5 \times 10^{-3}$ with
$\chi^2_{min}=37.4/30$ d.~o.~f.  We note that in contrast with the
earlier 504-day study of Ref.~\cite{bks} our results indicate that the
LMA solution is not only allowed, but actually slightly preferred. 
The existence of hints that the LMA MSW solution could be correct
was also discussed in Ref.\cite{bks99}.
On the other hand we find good quantitative agreement with the recent 
results of \cite{sk700} as well as qualitative agreement with the old results
of~\cite{Hata:1997di} based on a smaller sample. We show that seasonal
effects may be no-negligible in the lower part of the LMA region,
without conflict with the negative hints of a day-night variation.  In
particular the best-fit LMA solution predicts a seasonal effect of 8.5\%, 
6\% of which is due to the eccentricity of the Earth's orbit. 
For conversions into sterile neutrinos we find that only the SMA
solution is possible with best--fit point $\Delta m^2=5.0 \times
10^{-6}$ eV$^2$ and $\sin^2 2\theta=3.0 \times 10^{-3}$ and
$\chi^2_{min}=40.2/30$ d.~o.~f.  We also consider departures of the
Standard Solar Model (SSM) of Bahcall and Pinsonneault 1998 (BP98) by
allowing arbitrary $^8B$ and $hep$ fluxes. These modifications do not
affect significantly the oscillation parameters. We find that 
the best fit is obtained for $^8B/^8B_{SSM}=0.61$ and
$hep/hep_{SSM}=12$ for the SMA solution both for conversions into
active or sterile neutrinos and $^8B/^8B_{SSM}=1.37$ and
$hep/hep_{SSM}=38$ for the LMA solution.

\section{Data and Techniques}
\label{sec:data}

In order to study the possible values of neutrino masses and mixing
for the MSW solution to the solar neutrino problem, we have used data
on the total event rates measured at the Chlorine experiment in
Homestake \cite{homestake}, at the two Gallium experiments GALLEX and
SAGE \cite{gallex,sage} and at the water Cerenkov detector
Super--Kamiokande. Apart from total event rates we have in this case
the zenith angle distribution of the events, the electron recoil
energy spectrum and the seasonal distribution of events, all measured
with their recent 825-day data sample~\cite{sk700}.

We first describe our calculation of the different observables. 
For simplicity we consider the two-neutrino mixing case
\begin{equation}
\nu_e = \cos \theta ~\nu_1 + \sin \theta ~\nu_2 \; , \;
\nu_x = - \sin \theta ~\nu_1 + \cos \theta ~\nu_2 ~,
\label{eigendef}
\end{equation} 
where $x$ can label either an active, $x=\mu,\tau$, or sterile
neutrino, $x=s$.  In order to account for Earth regeneration effects,
we have determined the solar neutrino survival probability $P_{ee}$ in
the usual way, assuming that the neutrino state arriving at the Earth
is an incoherent mixture of the $\nu_1$ and $\nu_2$ mass eigenstates.
\begin{equation}
P_{ee} = P_{e1}^{Sun} P_{1e}^{Earth} + P_{e2}^{Sun} P_{2e}^{Earth}
\end{equation}
where $ P_{e1}^{Sun} $ is the probability that a solar neutrino, that
is created as $\nu_e $, leaves the Sun as a mass eigenstate $\nu _1$,
and $ P_{1e}^{Earth} $ is the probability that a neutrino which enters
the Earth as $\nu _1$ arrives at the detector as $\nu_e$. Similar
definitions apply to $P_{e2}^{Sun}$ and $P_{2e}^{Earth}$.

The quantity $P_{e1}^{Sun}$ is given, after discarding the fast
oscillating terms, as
\begin{equation}
P_{e1}^{Sun}  = 1- P_{e2}^{Sun}  = 
\frac{1}{2} + (\frac{1}{2} - P_{LZ})
cos[2\theta_m(t_0)]
\end{equation}
where $P_{LZ}$ denotes the improved Landau-Zener probability
\cite{LZ} and $\theta(t_0)_m$ is the mixing angle in matter
at the neutrino production point. In our calculations of the expected
event rates we have averaged this probability with respect to the
production point. The electron and neutron number density in the sun
and the production point distribution were taken from
Ref.~\cite{prod}.

In order to obtain the conversion probabilities in the Earth, 
$P_{ie}^{Earth}$, we integrate the evolution equation in matter
assuming a step function profile of the Earth matter density (the 
Earth as consisting of mantle and core  of constant densities equal to the
corresponding average densities, $\bar{\rho}_m\simeq 4.5$ $g/cm^3$ and 
$\bar{\rho}_c\simeq 11.5$ $g/cm^3$). To convert from the mass density
to electron and neutron number density we use the charge to nucleon ratio
$Z/A = 0.497$ for the mantle and $Z/A=0.467$ for the core.  
In the notation of Ref.~\cite{Akhmedov}, we obtain for  
$P_{2e}^{Earth}= 1-P_{1e}^{Earth}$
\begin{equation}
P_{2e}^{Earth}(\Phi)=
(Z sin \theta )^2 + (W_1cos \theta + W_3sin \theta )^2\\
\end{equation}
where $\theta$ is the mixing angle in vacuum and the Earth matter
effect is included in the formulas for $Z, W_1$ and $W_3$, 
which can be found in Ref.~\cite{Akhmedov}. $P_{2e}^{Earth}$ 
depends on the amount of Earth matter travelled by the neutrino 
on its way to the detector, or, in other words, on its arrival 
direction which is usually parametrized in terms of the nadir 
angle, $\Phi$, of the sun at the detector site. Due to this effect
the survival probability is, in general, time dependent.
This Earth regeneration effect is important in the study of the zenith 
angle distribution of events as well as in their seasonal variation 
\cite{ourseasonal,daynight}.   

\subsection{Rates}
\label{data:rates}

Here we update previous analyses of solar neutrino data
\cite{bks,Hata:1997di} by including the recent 825-day
Super--Kamiokande data sample. We perform a dedicated analysis of the
seasonal variation data and we anticipate its future role in
discriminating between different solutions of the solar neutrino
anomaly. Working in the context of the BP98 standard solar model of
Ref.~\cite{BP98} we also allow for a free normalization of the $^8B$
flux and of the hep flux. In our statistical treatment of the data we
follow closely the analysis of Ref.~\cite{fogli-lisi} with the 
updated uncertanties given in Refs.~\cite{BP98,prod}.

In our study we use the measured rates shown in Table~\ref{rates}.
For the combined fit we adopt the $\chi^2$ definition:
\begin{equation}
\chi^2_R=\sum_{i,j=1,3} (R^{th}_i- R^{exp}_i)
\sigma_{ij}^{-2} (R^{th}_j- R^{exp}_j)
\end{equation}
where $R^{th}_i$ is the theoretical prediction of the event rate in
detector $i$ and $R^{exp}_i$ is the measured rate. The error matrix
$\sigma_{ij}$ contains not only the theoretical uncertainties but also
the experimental errors, both systematic and statistical.

The general expression of the expected event rate in the presence of
oscillations in experiment $i$ is given by $R^{th}_i$ :
\begin{eqnarray}
R^{th}_i & = & \sum_{k=1,8} \phi_k
\int\! dE_\nu\, \lambda_k (E_\nu) \times 
\big[ \sigma_{e,i}(E_\nu)  \langle P_{ee} (E_\nu,t) \rangle \label{ratesth} \\
& &                            + \sigma_{x,i}(E_\nu) 
(1-\langle P_{ee} (E_\nu,t)\rangle )\big] \nonumber,
\end{eqnarray}  
where $E_\nu$ is the neutrino energy, $\phi_k$ is the total neutrino
flux and $\lambda_k$ is the neutrino energy spectrum (normalized to 1)
from the solar nuclear reaction $k$ ~\cite{Bspe} with the
normalization given in Ref.~\cite{BP98}. Here $\sigma_{e,i}$
($\sigma_{x,i}$) is the $\nu_e$ ($\nu_x$, $x=\mu,\,\tau$) interaction
cross section in the Standard Model~\cite{CrSe} with the target
corresponding to experiment $i$, and $\langle P_{ee} (E_\nu,t)
\rangle$ is the time--averaged $\nu_e$ survival probability.

For the Chlorine and Gallium experiments we use improved cross
sections $\sigma_{\alpha,i}(E)$ $(\alpha = e,\,x)$ from
Ref.~\cite{prod}. For the Super--Kamiokande experiment we calculate
the expected signal with the corrected cross section given in
Sec.~\ref{data:spec}.

The expected signal in the absence of oscillations, $R^{\rm BP98}_i$,
can be obtained from Eq.(\ref{ratesth}) by substituting $P_{ee}=1$. In
Table~\ref{rates} we also give the expected rates at the different
experiments which we obtain using the fluxes of Ref.~\cite{BP98}.

\subsection{Day-Night Variation}
\label{data:dn}

As already mentioned, in the MSW picture the expected event rates can
be different when the neutrinos travel through the Earth due to the
$\nu_e$ regeneration effect \cite{daynight}.  As a result in certain
regions of the oscillation parameters the expected event rates depend
on the zenith angle of the sun as observed from the experiment site,
since this determines the amount of Earth matter crossed by the
neutrino on its way to the detector.

The Super--Kamiokande Collaboration has studied the dependence of the
event rates with the period of time along the day and the night. 
They present their
results in the form of a zenith angle distribution of events.

In our analysis we have used the experimental results from the
Super--Kamiokande Collaboration on the zenith angle distribution of
events taken on 5 night periods and the day averaged value, shown in
Table~\ref{daynight} which we graphically reduced from
Ref.~\cite{sk700}

We define $\chi^2$ for the zenith angle data as:
\begin{equation}
\chi^2_Z=\sum_{i=1,6} 
\frac{(\alpha_{z} 
\frac{\displaystyle R^{th}_i}{\displaystyle R^{\rm BP98}_i}
- R^{exp}_i)^2}{\sigma_i^2}
\end{equation}
where we have neglected the possible correlation between the errors of
the different angular bins which could arise from systematical
uncertainties. The factor $\alpha_{z}$ is included in order to avoid
over-counting the data on the total event rate which is already
included in $\chi^2_R$.

We compute the expected event rate in the period $i$ in the presence 
of oscillations as,
\begin{eqnarray} 
R^{th}_i & = & \frac{\displaystyle 1}{\displaystyle\Delta \tau_i}
\int_{\tau(cos\Phi_{min,i})}^{\tau(cos\Phi_{max,i})}  d\tau
\sum_{k=1,8} \phi_k\int\! dE_\nu\, \lambda_k (E_\nu) \times 
\big[ \sigma_{e,i}(E_\nu) \langle P_{ee} (E_\nu,\tau) \rangle  
\label{eq:daynight}\\ 
& &+ \sigma_{x,i}(E_\nu) 
(1-\langle P_{ee} (E_\nu,\tau)\rangle )\big] \nonumber
\end{eqnarray}  
where $\tau$ measures the yearly averaged length of the period $i$ 
normalized to 1, so $\Delta\tau_i=\tau(cos\Phi_{max,i})-\tau
(cos\Phi_{min,i})=$
.500, .086, .091, .113, .111, .099 for the day and five night periods 
respectively.

Super--Kamiokande has also presented their results on the day-night
variation in the form of a day-night asymmetry,
\begin{equation}
A_{D/N}=\frac{Day-Night}{Day+Night}=
-0.065\pm 0.031 \mbox{(stat.)}\pm 0.013\mbox{(syst.)}
\end{equation}
Since the information included in the zenith angle dependence already
contains the day-night asymmetry, we have not added the asymmetry as
an independent observable in our fit. Notice also, that being a ratio
of event rates, the asymmetry is not a Gaussian-distributed observable
and therefore should not be included in a $\chi^2$ analysis.

\subsection{Recoil Electron Spectrum}
\label{data:spec}

The Super-Kamiokande Collaboration has also measured the recoil
electron energy spectrum.  In their published analysis \cite{skspec}
after 504 days of operation they present their results for energies
above 6.5 MeV using the Low Energy (LE) analysis in which the recoil
energy spectrum is divided into 16 bins, 15 bins of 0.5 MeV energy
width and the last bin containing all events with energy in the range
14 MeV to 20 MeV.  Below 6.5 MeV the background of the LE analysis
increases very fast as the energy decreases. Super--Kamiokande has
designed a new Super Low Energy (SLE) analysis in order to reject this
background more efficiently so as to be able to lower their threshold
down to 5.5 MeV. In their 825-day data \cite{sk700} they have used the
SLE method and they present results for two additional bins with
energies between 5.5 MeV and 6.5 MeV.

In our study we use the experimental results from the
Super--Kamiokande Collaboration on the recoil electron spectrum on the
18 energy bins including the results from the LE analysis for the 16
bins above 6.5 MeV and the results from the SLE analysis for the two
low energy bins below 6.5 MeV, shown in Table~\ref{spectrum}.

Notice that in Table~\ref{spectrum} we have symmetrized the errors to
be included in our $\chi^2$ analysis. We have explicitly checked that
the exclusion region is very insensitive to this symmetrization.  We
define $\chi^2$ for the spectrum as
\begin{equation}
\chi^2_S=\sum_{i,j=1,18} 
(\alpha_{sp}\frac{\displaystyle R^{th}_i}{R^{\rm BP98}_i} 
-R^{exp}_i)
\sigma_{ij}^{-2} (\alpha_{sp}\frac{\displaystyle R^{th}_j}{R^{\rm BP98}_i}
 - R^{exp}_j)
\end{equation}
where
\begin{equation}
\sigma^2_{ij}=\delta_{ij}(\sigma^2_{i,stat}+\sigma^2_{i,uncorr})+
\sigma_{i,exp} \sigma_{j,exp}+\sigma_{i,cal}\sigma_{j,cal}
\end{equation}
Again, we introduce a normalization factor $\alpha_{sp}$ in order to
avoid double-counting with the data on the total event rate which is
already included in $\chi^2_R$.  Notice that in our definition of
$\chi^2_S$ we introduce the correlations amongst the different
systematic errors in the form of a non-diagonal error matrix in
analogy to our previous analysis of the total rates.  These
correlations take into account the systematic uncertainties related to
the absolute energy scale and energy resolution, which were not yet
available at the time the analysis of Ref.~\cite{bks} was performed.
Note that our procedure is different from that used by the
Super-Kamiokande collaboration. However we will see in
Sec.~\ref{fit:spec} that both methods give very similar results for
the exclusion regions. This provides a good test of the robustness of
the results of the fits, in the sense that they do not depend on the
details of the statistical analysis.

The general expression of the expected rate in the presence of
oscillations $R^{th}$ in a bin, is given from Eq.(\ref{ratesth}) but
integrating within the corresponding electron recoil energy bin and
taking into account that the finite energy resolution implies that the
{\em measured } kinetic energy $T$ of the scattered electron is
distributed around the {\em true } kinetic energy $T'$ according to a
resolution function $Res(T,\,T')$ of the form~\cite{Kr97}:
\begin{equation}
Res(T,\,T') = \frac{1}{\sqrt{2\pi}s}\exp\left[
{-\frac{(T-T')^2}{2 s^2}}\right]\ ,
\end{equation}
where
\begin{equation}
s = s_0\sqrt{T'/{\rm MeV}}\ ,
\label{Delta}
\end{equation}
and $s_0=0.47$ MeV for Super--Kamiokande \cite{sk1,Faid}. On the other
hand, the distribution of the true kinetic energy $T'$ for an
interacting neutrino of energy $E_\nu$ is dictated by the differential
cross section $d\sigma_\alpha(E_\nu,\,T')/dT'$, that we take from
\cite{CrSe}. The kinematic limits are:
\begin{equation}
0\leq T' \leq {\overline T}'(E_\nu)\ , 
\ \ {\overline T}'(E_\nu)=\frac{E_\nu}{1+m_e/2E_\nu}\ .
\end{equation}
For assigned values of $s_0$, $T_{\rm min}$, and $T_{\rm max}$, the
corrected cross section $\sigma_{\alpha}(E)$ $(\alpha = e,\,x)$ is
given as
\begin{equation}
\sigma_{\alpha}(E_\nu)=\int_{T_{\rm min}}^{T_{\rm max}}\!dT
\int_0^{{\overline T}'(E_\nu)}
\!dT'\,Res(T,\,T')\,\frac{d\sigma_{\alpha}(E_\nu,\,T')}{dT'}\ .
\label{sigma}
\end{equation}
In Fig.~\ref{spectrum} we show our results for the recoil electron
spectrum in the absence of oscillations and compare it with the 
expectations from the Super--Kamiokande Monte-Carlo. We see that 
the agreement is excellent. In this figure no normalization has been 
included. 

\subsection{Seasonal Variation} 
\label{data:sea}

After 825 days of operation, Super--Kamiokande has also presented
preliminary results on the seasonal variation of the neutrino event
rates \cite{sk700} which seem to hint at a seasonal variation of the
data, especially for recoil electron energies above 11.5 MeV (see
Table ~\ref{seasonal}).  As discussed in Ref.~\cite{ourseasonal,v99}
the expected MSW event rates do exhibit a seasonal effect.  In the LMA
region such dependence can be expected mainly due to the different
night duration throughout the year at the experimental site and also
due to the different averaged Earth densities crossed by the neutrino
during the night periods, which lead to a seasonal-dependent $\nu_e$
regeneration effect in the Earth.  On the other hand, in the SMA
region, Earth matter effects are due to resonant conversion of
neutrinos in the Earth which is only possible when neutrinos travel
both through the mantle and the core~\cite{daynight}.  Thus we find
that in the SMA region the seasonal variation is associated with the
fact that at Super--Kamiokande site only in October-March nights there
are neutrinos arriving at sufficiently low zenith angle to satisfy the
resonant condition.

We define  $\chi^2$ for the seasonal variation data as,
\begin{equation}
\chi^2_{Sea}=\sum_{i=1,8} 
\frac{(\alpha_{sea} 
\frac{\displaystyle
R^{th}_i}{R^{\rm BP98}_i \displaystyle}- R^{exp}_i)^2}{\sigma_i^2}
\end{equation}
where, as before, the normalization factor $\alpha_{sea}$ is
introduced to avoid double-counting.

Taking into account the relative position of the Super--Kamiokande
setup in each period of the year, we calculate the distribution of the
events as
\begin{equation}
\frac{R^{th}_i(t_i, \Delta t)}{R^{\rm BP98}_i(\Delta t)}=
\frac
{\displaystyle \int_{t_i-\Delta t/2}^{t_i+\Delta t/2} dt\,
R^{th}(t)}{\displaystyle \Delta t R^{th}(t)} 
\end{equation}
Here $\Delta t=1.5$ months and $R^{th}(t)$ is obtained from
Eq.(\ref{ratesth}) but using the time--dependent survival
probabilities and integrating the recoil electron energy above 11.5
MeV.  Notice that, unlike in the 708 days data sample, in order to
compare our results with the recent data on the seasonal dependence of
the event rates from the Super--Kamiokande Collaboration for the 825
data sample, we have not included the geometrical seasonal neutrino
flux variation due to the variation of the Sun-Earth distance arising
from the Earth's orbit eccentricity because the new Super--Kamiokande
data is already corrected for this geometrical variation.

\section{Fits: Results}
\label{sec:fits}

We now turn to the results of our fits with the observables described
above. We have obtained the regions of allowed oscillation parameters
$\Delta m^2$-$\sin^2 2\theta$ by obtaining the minimum $\chi^2$ and
imposing the condition $\chi^2\leq\chi^2_{min}$+$\Delta\chi^2 (2,CL)$
where $\Delta\chi^2(2,CL)= 4.61 (9.21)$ for 90\% (99\%) CL regions.

\subsection{Rates}
\label{fit:rates}

We first determine the allowed range of oscillation parameters using
only the total event rates of the Chlorine, Gallium and
Super--Kamiokande experiments. The average event rates for these
experiments are summarized in Table~\ref{rates}. We have not included
in our analysis the Kamiokande data \cite{kamioka} as it is well in
agreement with the results from the Super--Kamiokande experiment and
the precision of this last one is much higher \cite{sk700}. For the
Gallium experiments we have used the weighted average of the results
from GALLEX \cite{gallex} and SAGE \cite{sage} detectors.

Using the predicted fluxes from the BP98 model the $\chi^2$ for the
total event rates is $\chi^2_{SSM}=62.4$ for 3 d.~o.~f.
This means that the SSM together with the SM of particle interactions
can explain the observed data with a probability lower than
$10^{-12}$!

In the case of active-active neutrino oscillations we find that the
best--fit point is obtained for the SMA solution with
\begin{eqnarray}
\Delta m^2 & =&  5.6\times 10^{-6} \mbox{eV}^2\; , \nonumber \\ 
\sin^2 2\theta& = & 6.3\times 10^{-3}\; , \label{smarates} \\
\chi^2_{min}& = & 0.37  \;,\nonumber
\end{eqnarray}
what implies that the solution is acceptable with a 55\% CL.
 
There are two more local minima of $\chi^2$. One is the LMA solution
with the best fit for
\begin{eqnarray}
\Delta m^2 & = & 1.4\times 10^{-5} \mbox{eV}^2\; , \nonumber \\ 
\sin^2 2\theta & = & 0.67\;, \label{lmarates} \\
\chi^2_{min}& = & 2.92 \; , \nonumber 
\end{eqnarray}
which is acceptable with at 91\% CL. The other is the LOW solution
with best--fit point
\begin{eqnarray}
\Delta m^2 & = & 1.3\times 10^{-7} \mbox{eV}^2\; , \nonumber \\ 
\sin^2 2\theta & = & 0.94\;,  \label{lowrates} \\
\chi^2_{min}& = & 7.4 \; , \nonumber 
\end{eqnarray}
which is only acceptable at 99\% CL. 

In the case of active-sterile neutrino oscillations the best--fit
point is obtained for the SMA solution with
\begin{eqnarray}
\Delta m^2 & =&  5.0\times 10^{-6} \mbox{eV}^2\; ,\nonumber \\
\sin^2 2\theta& = &5.0\times 10^{-3}\; ,\label{strates}\\
\chi^2_{min}& = & 2.6  \;,\nonumber
\end{eqnarray}
acceptable at 90\% CL. The LMA and LOW solutions are not acceptable
for oscillation into sterile neutrinos. In those regions
$\chi^2_{min}\geq 19.5$ implying that they are excluded at the
99.999\% CL.  Unlike active neutrinos which lead to events in the
Super--Kamiokande detector by interacting via neutral current with the
electrons, sterile neutrinos do not contribute to the Super--Kamiokande
event rates.  Therefore a larger survival probability for $^8B$
neutrinos is needed to accommodate the measured rate. As a consequence
a larger contribution from $^8B$ neutrinos to the Chlorine and Gallium
experiments is expected, so that the small measured rate in Chlorine
can only be accommodated if no $^7Be$ neutrinos are present in the
flux. This is only possible in the SMA solution region, since in the
LMA and LOW regions the suppression of $^7Be$ neutrinos is not enough.

\begin{figure}
\begin{center}
\mbox{\epsfig{file=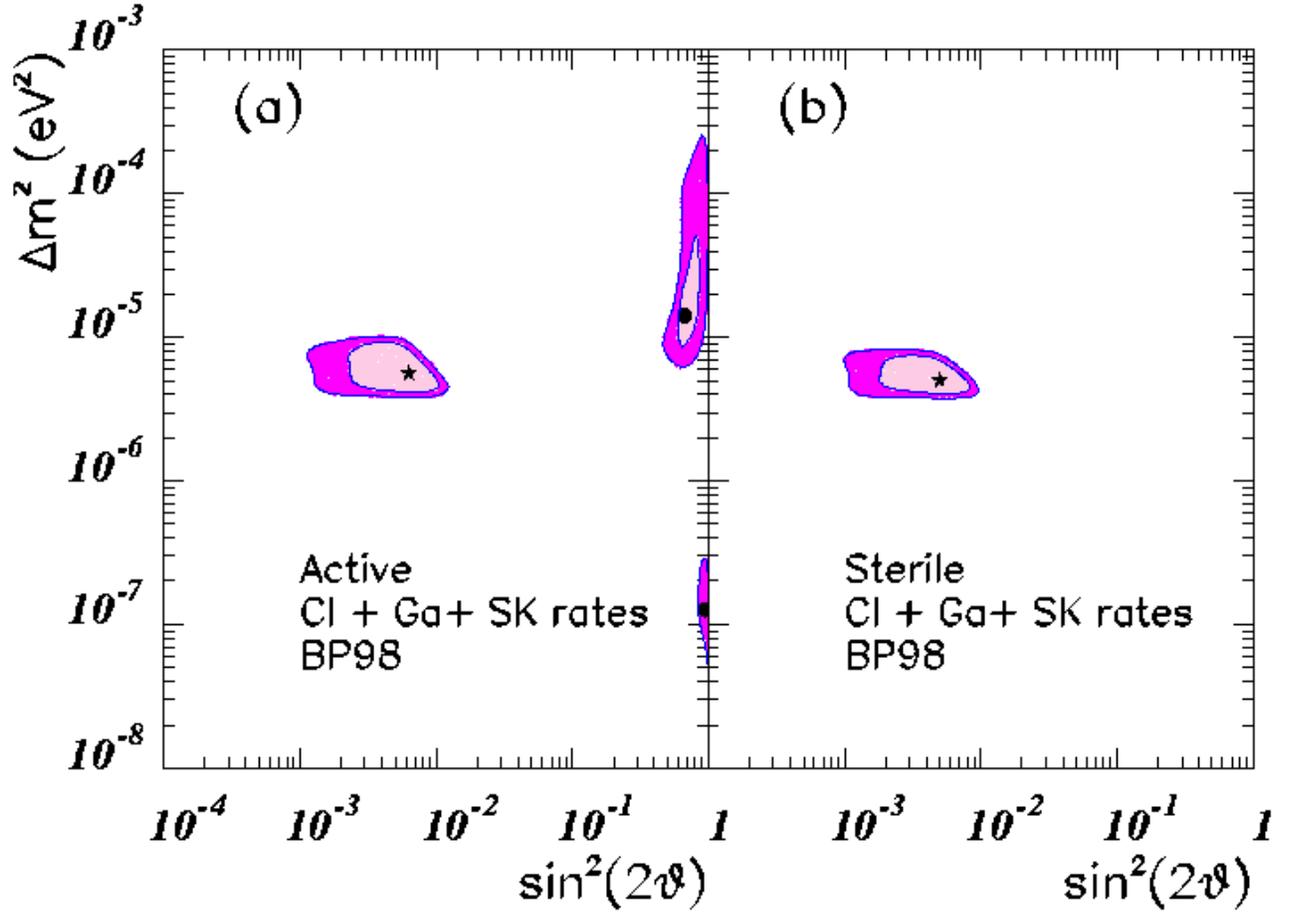,height=13cm}}
\end{center}
\caption{Allowed regions in  $\Delta m^2$ and $\sin^2\theta$ 
from the measurements of the total event rates at Chlorine, Gallium
and Super--Kamiokande (825-day data sample) for active-active
transitions {\bf(a)} and active-sterile {\bf(b)}.  The darker
(lighter) areas indicate the 99\% (90\%)CL regions. Global best--fit point
is indicated by a star. Local best--fit points are indicated by a dot.}
\label{fig:rates}
\end{figure}

In Fig.~\ref{fig:rates} we show the 90\% and 99\% CL allowed regions
in the plane $\Delta m^2$-$\sin^2 2\theta$. The best--fit points in
each region are marked. We find that as far as the analysis
of the total rates is concerned, there is no substantial change in the
best--fit points in the three regions as compared to the previous most
recent analysis including the 504 days of Super--Kamiokande data
\cite{bks}.

\begin{figure}
\begin{center}
\mbox{\epsfig{file=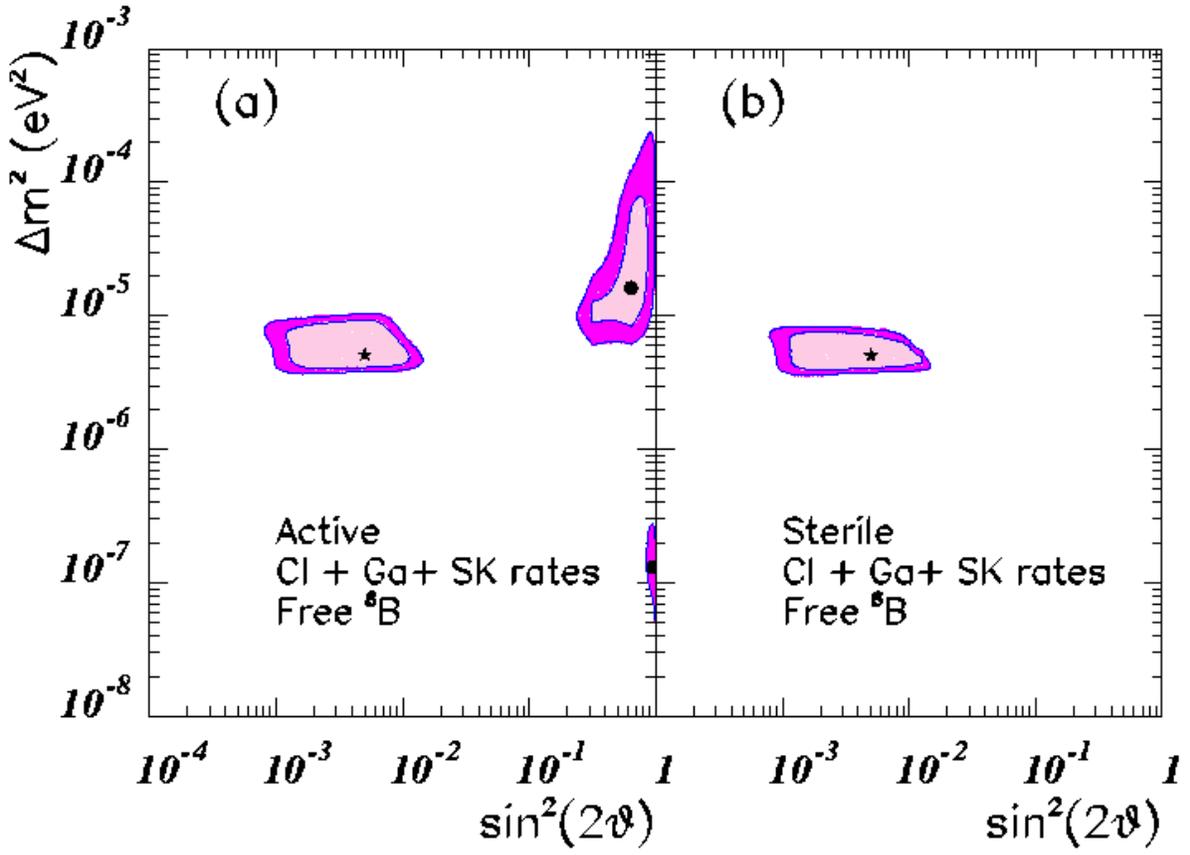,height=13cm}}
\end{center}
\caption{Same as previous figure but allowing a free $^8B$ flux
normalization.}
\label{fig:ratesbo}
\end{figure}

We have also considered the possibility of departing from the SSM of
BP98 by allowing a free normalization of the $^8B$ flux and we treat
this normalization as a free parameter $\beta$ in our analysis. 
Figure \ref{fig:ratesbo} shows the allowed regions in the MSW
parameter space when $\beta$ is allowed to take arbitrary values.  The
best fit SMA solution is obtained for
\begin{eqnarray}
\Delta m^2 & =&  5.0\times 10^{-6} \mbox{eV}^2\; ,\nonumber \\ 
\sin^2 2\theta& = & 5.0\times 10^{-3}\; , \\
\beta&=&0.82  \nonumber \\
\chi^2_{min}& = & 0.05  \;.\nonumber
\end{eqnarray}
The best fit for the LMA solution occurs at: 
\begin{eqnarray}
\Delta m^2 & = & 1.6\times 10^{-5} \mbox{eV}^2\; , \nonumber \\
\sin^2 2\theta & = & 0.63\;,  \\
\beta&=&1.32 \nonumber \\
\chi^2_{min}& = & 0.47 \; , \nonumber 
\end{eqnarray}
and the LOW solution has its best--fit point at $\beta=.98$ and 
therefore coincides with the one obtained in Eq.(\ref{lowrates}).

In the case of active-sterile neutrino oscillations the best--fit
point is obtained for the SMA solution :
\begin{eqnarray}
\Delta m^2 & =&  5.0\times 10^{-6} \mbox{eV}^2\; ,\nonumber \\ 
\sin^2 2\theta& = & 3.2\times 10^{-3}\; , \\
\beta&=&0.75  \nonumber \\
\chi^2_{min}& = & 2.16  \;.\nonumber
\end{eqnarray}

For all solutions we find that the main effect of considering a free
normalization of the $^8B$ flux is upon the quality of the fits, as
measured by the depth of the $\chi^2$. Next comes the position of the
best--fit points, mainly a reduction in the value of the mixing angle.
The allowed regions are considerably enlarged as can be seen by 
comparing Figs.~\ref{fig:rates} and \ref{fig:ratesbo}.

\subsection{Yearly Averaged Zenith Angle Dependence}
\label{fit:dn}

We now study the constraints on the oscillation parameters from the
Super--Kamiokande Collaboration measurement of the zenith angle
distribution of events. We use the data taken on 5 night periods and
the day averaged value shown in Table~\ref{daynight} which we
graphically reduced from Ref.~\cite{sk700}.
 
\begin{figure}
\begin{center}
\mbox{\epsfig{file=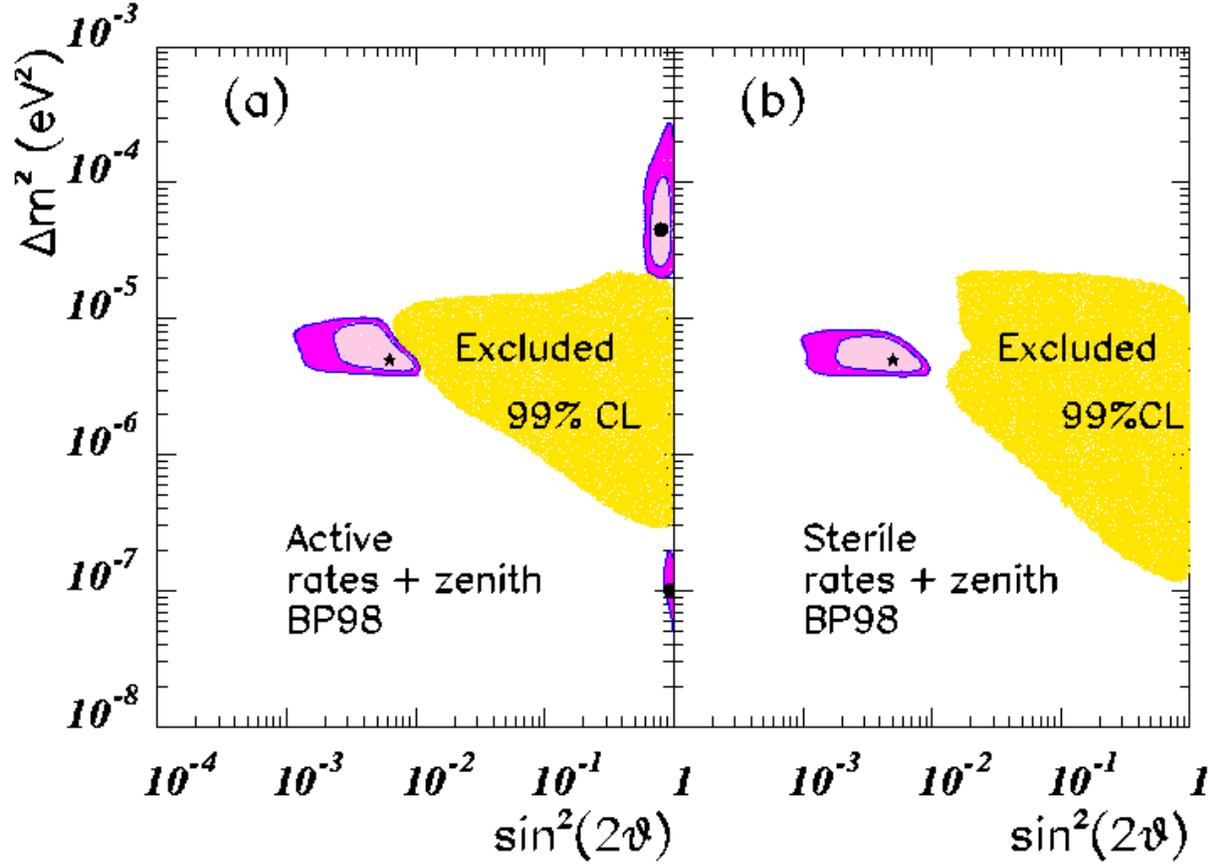,height=13cm}}
\end{center}
\caption{Allowed regions in  $\Delta m^2$ and $\sin^2\theta$ 
from the measurements of the total event rates at Chlorine, Gallium
and Super--Kamiokande (825-day data sample) combined with the zenith
angle distribution observed in Super--Kamiokande for active-active
{\bf(a)} and active-sterile transitions {\bf(b)}. The darker (lighter)
areas indicate 99\% (90\%)CL regions.  Global best--fit point
is indicated by a star. Local best--fit points are indicated by a dot.
The shadowed area represents the
region excluded by the zenith angle distribution data at 99\% CL.}
\label{fig:ratesdn}
\end{figure}

Considering only the zenith angle data (not including the information
from the total rates), the best--fit point in the case of
active-active neutrino oscillations is obtained for $\Delta m^2 =
4.5\times 10^{-5}$~eV$^2$ and $\sin^2 2\theta=1.0$ with
$\chi^2_{min}=2.3$ for 3 d.~o.~f. and in the case of active-sterile
neutrino oscillations is obtained for $\Delta m^2 = 3.2\times
10^{-5}$~eV$^2$ and $\sin^2 2\theta=.98$ with $\chi^2_{min}=2.2$. With
these values we calculate the excluded region of parameters at the
99\% CL, shown in Fig \ref{fig:ratesdn}.  Notice that the zenith angle
data favours the LMA solution of the solar neutrino problem.

When combining the information from both total rates and zenith angle
data, we obtain that in the case of active-active neutrino
oscillations the best--fit point is still obtained for the SMA
solution with
\begin{eqnarray}
\Delta m^2 & =&  5.0\times 10^{-6} \mbox{eV}^2\; , \nonumber \\ 
\sin^2 2\theta& = & 6.3\times 10^{-3}\; , \label{smardn} \\
\chi^2_{min}& = & 5.9  \;,\nonumber
\end{eqnarray}
for 6 d.~o.~f. which is acceptable with a 56\% CL.  However the LMA
solution becomes relatively better with a local minimum at
\begin{eqnarray}
\Delta m^2 & = & 4.5\times 10^{-5} \mbox{eV}^2\; , \nonumber \\ 
\sin^2 2\theta & = & 0.80\;, \label{lmardn} \\
\chi^2_{min}& = & 7.2 \; , \nonumber 
\end{eqnarray}
valid at 70\% CL. This difference arises from the fact that, although
small, some effect is observed in the zenith angle dependence which
points towards a larger event rate during the night than during the
day, and that this difference is constant during the night as expected
for the LMA solution~\cite{bks99}. In the SMA solution, however, the
enhancement is expected to occur mainly in the fifth night
\cite{daynight}.
  
The LOW solution is almost un--modified and presents the best--fit
point at
\begin{eqnarray}
\Delta m^2 & = & 1.0\times 10^{-7} \mbox{eV}^2\; , \nonumber \\ 
\sin^2 2\theta & = & 0.94\;,  \label{lowrdn} \\
\chi^2_{min}& = & 12.7 \; , \nonumber 
\end{eqnarray}
which is acceptable at 95\% CL. 

In the case of active-sterile neutrino oscillations the best--fit point is 
obtained  for the SMA solution with  
\begin{eqnarray}
\Delta m^2 & =&  5.0\times 10^{-6} \mbox{eV}^2\; ,\nonumber \\
\sin^2 2\theta& = & 5.0\times 10^{-3}\; ,\label{strdn}\\
\chi^2_{min}& = & 8.1 \;,\nonumber
\end{eqnarray}
valid at 77\% CL. 

Figure \ref{fig:ratesdn} shows the regions excluded at 99\% CL by the
zenith angle data alone, together with the 90\% and 99\% CL allowed
regions in the plane $\Delta m^2$-$\sin^2 2\theta$ from the combined
analysis of rates plus zenith angle data. The best--fit points in each
region are indicated.  The main difference with respect to
Fig.~\ref{fig:rates} is observed in the LMA allowed region which is
cut from below, as the expected day-night variation is too large for
smaller neutrino mass differences. 

\subsection{Recoil Electron Spectrum}
\label{fit:spec}
 
We now present the results of the study of the recoil electron
spectrum data observed in Super--Kamiokande. Using the method
described in Sec.~\ref{data:spec} we obtain that the $\chi^2$ for the
undistorted energy spectrum (Standard Model case) is 20.1 for 17
d.~o.~f..  This corresponds to an agreement of the measured with the
expected energy shape at the 27\% CL. This value depends on the
degree of correlation allowed between the different errors. In this
way, reducing the correlation in the error matrix the agreement
decreases down to 13\%.  This suggests that the correlations amongst
the different systematic uncertainties related to the energy
resolution will play an important role in the analysis of the energy
spectrum.  Indeed in our definition of $\chi^2_S$ we introduce the
correlations amongst the different systematic errors in the form of a
non-diagonal error matrix in analogy to our previous analysis of the
total rates. These correlations take into account the systematic
uncertainties related to the energy resolution. We believe that these
might constitute the main difference between our treatment of the spectrum
data and the earlier one presented in Ref.~\cite{bks}, when this
experimental information was still unavailable.

\begin{figure}
\begin{center}
\mbox{\epsfig{file=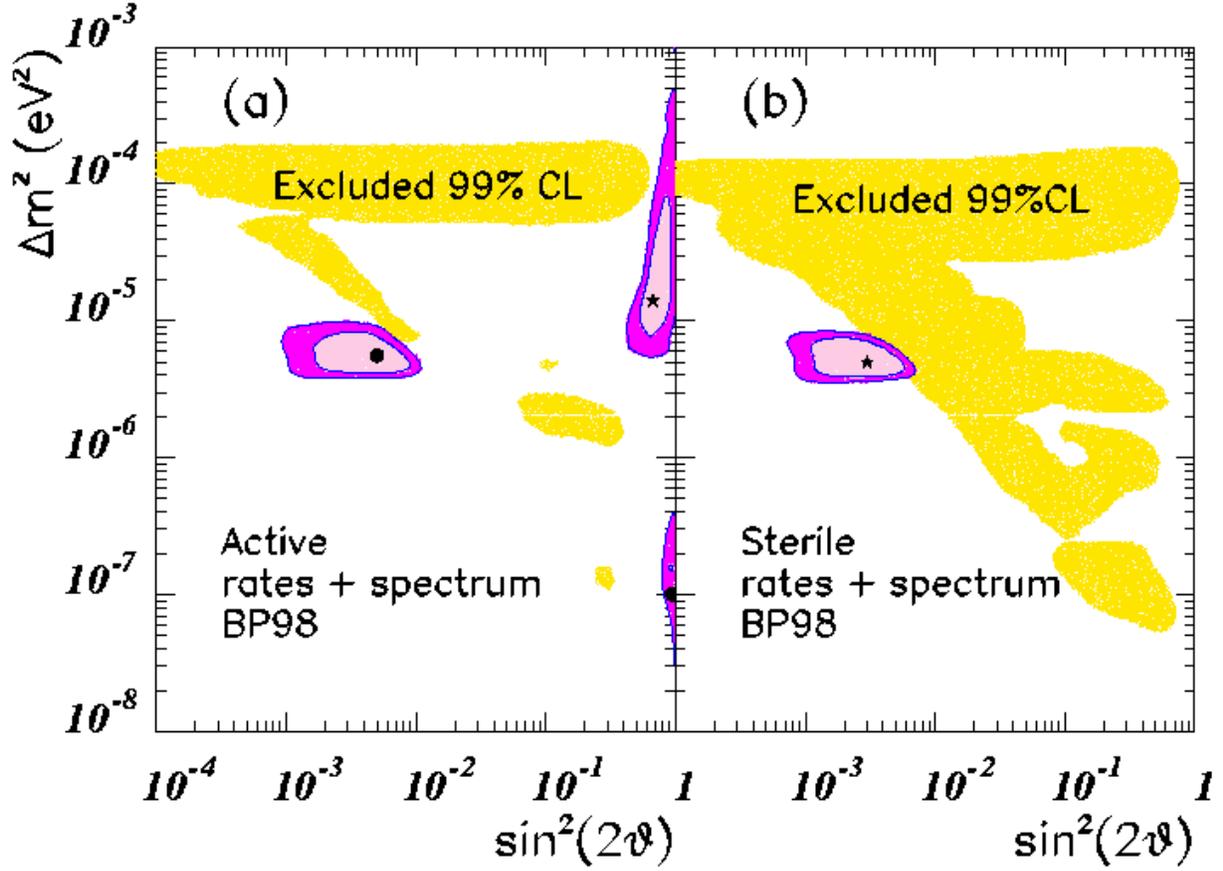,height=13cm}}
\end{center}
\vglue -.5cm
\caption{
Allowed regions in $\Delta m^2$ and $\sin^2\theta$ from the
measurements of the total event rates at the Chlorine, Gallium and
Super--Kamiokande (825-day data sample) combined with the recoil
electron spectrum data observed in Super--Kamiokande for active-active
oscillations {\bf(a)} and active-sterile oscillations {\bf(b)} . The
darker (lighter) area indicate 99\% (90\%)CL regions.  
Global best--fit point
is indicated by a star. Local best--fit points are indicated by a dot.
The shadowed area
represents the region excluded by the spectrum data at (99\%)CL.}
\label{fig:ratesspec}
\end{figure}

We find that the best fit to the spectrum in the MSW plane for the
case of active-active neutrino oscillations is obtained for $\Delta
m^2 = 6.3\times 10^{-6}$ eV$^2$ and $\sin^2 2\theta=0.08$ with
$\chi^2_{min}=17.9$. For the case of active-sterile neutrino
oscillations we get $\Delta m^2 = 6.3\times 10^{-6}$ eV$^2$ and
$\sin^2 2\theta=.08$ with $\chi^2_{min}=17$. With these values we
obtain the excluded region of parameters at the 99\% CL shown in
Fig.~\ref{fig:ratesspec}.

We see that our results for the exclusion regions are quantitatively
very similar to those of the Super-Kamiokande collaboration
\cite{sk700}, even though our procedure is different from
theirs. This provides a good test of the robustness of the results of
the fit of the spectrum, in the sense that they do not depend on the
details of the statistical analysis. 
In contrast we note that our results are different from those of
Ref.~\cite{bks}.

When combining the information from both total rates and the recoil
energy spectrum data we obtain that in the case of active-active
neutrino oscillations both SMA and LMA solutions lead to fits to the
data if similar quality. In this way, the best--fit for the LMA
solution is
\begin{eqnarray}
\Delta m^2 & = & 1.4\times 10^{-5} \mbox{eV}^2\; , \nonumber \\ 
\sin^2 2\theta & = & 0.67\;, \label{lmarspec} \\
\chi^2_{min}& = & 22.5 \; , \nonumber 
\end{eqnarray}
while for the SMA solution we find
\begin{eqnarray}
\Delta m^2 & =&  5.6\times 10^{-6} \mbox{eV}^2\; , \nonumber \\ 
\sin^2 2\theta& = & 5.0\times 10^{-3}\; , \label{smarspec} \\
\chi^2_{min}& = & 23.4  \;,\nonumber
\end{eqnarray}
for 18. d.~o.~f. which are acceptable at ~83\%. Finally for the LOW
solution we find
\begin{eqnarray}
\Delta m^2 & = & 1.0\times 10^{-7} \mbox{eV}^2\; , \nonumber \\ 
\sin^2 2\theta & = & 0.94\;, \label{lowrspec} \\
\chi^2_{min}& = & 26.7 \; , \nonumber 
\end{eqnarray}
with agreement with data only at the 8.5\% CL. 

In the case of active-sterile neutrino oscillations the best--fit point is 
obtained for the SMA solution with  
\begin{eqnarray}
\Delta m^2 & =&  5.0\times 10^{-6} \mbox{eV}^2\; ,\nonumber \\
\sin^2 2\theta& = & 3.0\times 10^{-3}\; ,\label{strspec}\\
\chi^2_{min}& = & 26.3 \;.\nonumber
\end{eqnarray}
acceptable at ~90\%.

In Fig.~\ref{fig:ratesspec} we plot the excluded region at 99\% CL by
the energy spectrum data together with the 90\% and 99\% CL allowed
regions in the plane $\Delta m^2$-$\sin^2 2\theta$ from the combined
analysis. The best--fit points in each region are marked.

The main point here is that the oscillation hypothesis does not improve
considerably the fit to the energy spectrum as compared to the
no-oscillation hypothesis. In this connection it has been suggested
\cite{bk98} that better description can be obtained by allowing a
larger flux of hep neutrinos as they contribute mainly to the end part
of the spectrum. In order to account for this possibility we have also
analysed the data allowing for a free normalization of the $^8B$ and
$hep$ fluxes, treating them as a free parameters $\beta$ and $\gamma$
correspondingly. 
When doing so we find that the no-oscillation
hypothesis gives $\chi^2$ = 17.4 for 15 d.~o.~f. for $\beta=0.45$ and
$\gamma=13.5$.

When combining the information from  both total rates and the 
recoil energy spectrum data in the case of active-active 
neutrino oscillations we obtain that for the LMA:
\begin{eqnarray}
\Delta m^2 & = & 1.6\times 10^{-5} \mbox{eV}^2\; , \nonumber \\ 
\sin^2 2\theta & = & 0.63\;, \label{lmarspecbh} \\
\beta=1.3 & & \gamma= 33  \nonumber \\
\chi^2_{min}& = & 17.5\; , \nonumber 
\end{eqnarray}
for 16 d.~o.~f. which is acceptable at ~66\% while for the SMA solution 
\begin{eqnarray}
\Delta m^2 & =&  5.0\times 10^{-6} \mbox{eV}^2\; , \nonumber \\ 
\sin^2 2\theta& = & 2.5\times 10^{-3}\; , \label{smarepecbh} \\
\beta=.61 & & \gamma= 13  \nonumber \\
\chi^2_{min}& = & 19.5  \;,\nonumber
\end{eqnarray}
which is acceptable at ~75\% CL. Finally for the LOW solution we find
\begin{eqnarray}
\Delta m^2 & = & 1.0\times 10^{-7} \mbox{eV}^2\; , \nonumber \\ 
\sin^2 2\theta & = & 0.94\;, \label{lowrspecbh} \\
\beta=.97 & & \gamma= 22  \nonumber \\
\chi^2_{min}& = & 25\; , \nonumber 
\end{eqnarray}
acceptable at ~93\%.

In the case of active-sterile neutrino oscillations the best--fit point is 
obtained  for the SMA solution :
\begin{eqnarray}
\Delta m^2 & =&  5.0\times 10^{-6} \mbox{eV}^2\; ,\nonumber \\ 
\sin^2 2\theta& = & 2.0\times 10^{-3}\; , \\
\beta=0.61 & & \gamma= 12 \nonumber \\
\chi^2_{min}& = & 22  \;.\nonumber
\end{eqnarray}
which is acceptable at ~89\% CL. 

\begin{figure}
\begin{center}
\mbox{\epsfig{file=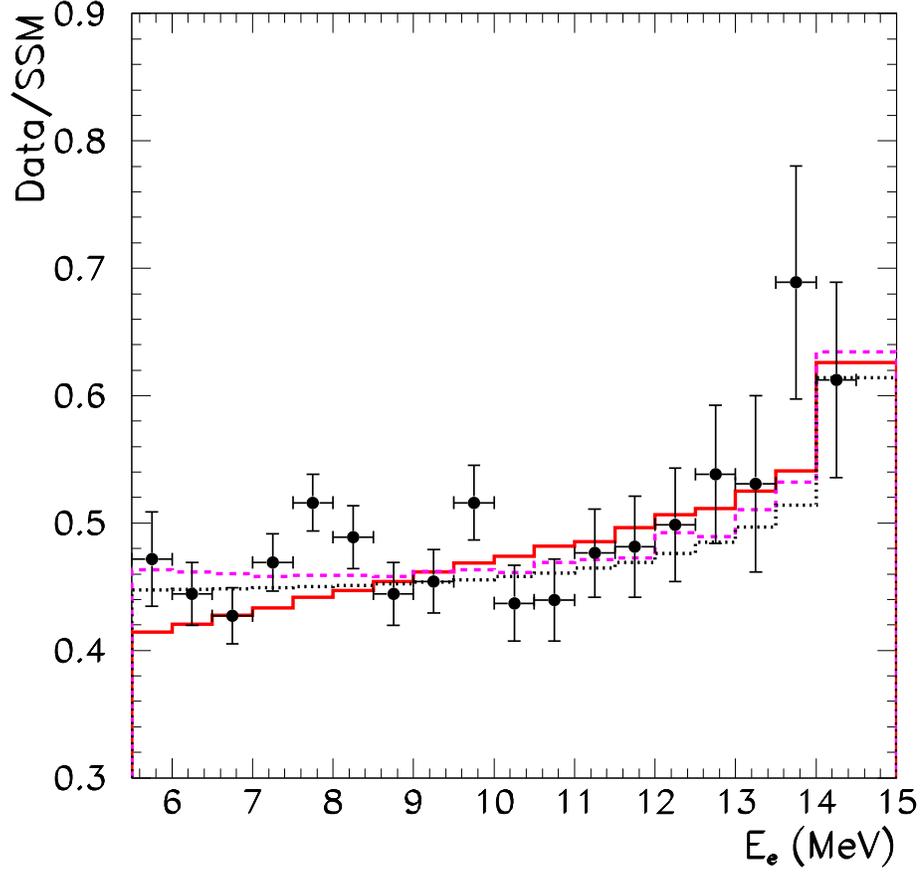,height=13cm}}
\end{center}
\vglue -.5cm
\caption{Expected normalized recoil electron energy spectrum 
compared with the experimental data~\protect\cite{sk700}. The solid
line represents the prediction for the best--fit SMA solution with
free $^8B$ and $hep$ normalizations ($\beta = 0.61$, $\gamma=12$),
while the dashed line gives the corresponding prediction for the
best--fit LMA solution ($\beta = 1.37$, $\gamma=38$). Finally, the
dotted line represents the prediction for the best non-oscillation
scheme with free $^8B$ and $hep$ normalizations ($\beta=0.45$,
$\gamma=13.5$).}
\label{fig:specnorm}
\end{figure}

In Fig.~\ref{fig:specnorm} we display the normalized expected energy
spectra for SMA, LMA solutions for active-active oscillations and for
no-oscillation with non-standard $^8B$ and $hep$ fluxes.

\subsection{Seasonal Dependence} 
\label{fit:sea}

Recently Super--Kamiokande has also presented preliminary results
which seem to hint for a seasonal variation of the event rates,
especially for recoil electron energy above 11.5 MeV. 
As explained in Sec. \ref{data:sea} the
expected MSW event rates do exhibit a seasonal effect due to the
the seasonal-dependent $\nu_e$ regeneration effect in the Earth.
We explore here the constraints on the MSW oscillation parameters which
can be extracted from the seasonal dependence data given in Table
~\ref{seasonal}.

Considering only the seasonal variation data above 11.5 MeV (allowing
a free normalization for the $^8B$ flux) we find that the the SSM
yields a value of $\chi^2=8. $ for 7 d.~o.~f.  This shows that the data is
still not precise enough to enable one to draw any definitive
conclusion.  However one can still obtain some preliminary information
on the MSW parameters from the analysis of these data. In this
way, when allowing oscillations into active flavours we obtain the
best--fit point for $\Delta m^2 = 3.2\times 10^{-6}$~eV$^2$ and
$\sin^2 2\theta=.1$ with $\chi^2_{min}=1.8$ for 5 d.~o.~f. and in the
case of active-sterile neutrino oscillations best fit is obtained for
$\Delta m^2 = 1.3\times 10^{-6}$~eV$^2$ and $\sin^2 2\theta=0.1$ with
$\chi^2_{min}=2.0$.

\begin{figure}
\begin{center}
\mbox{\epsfig{file=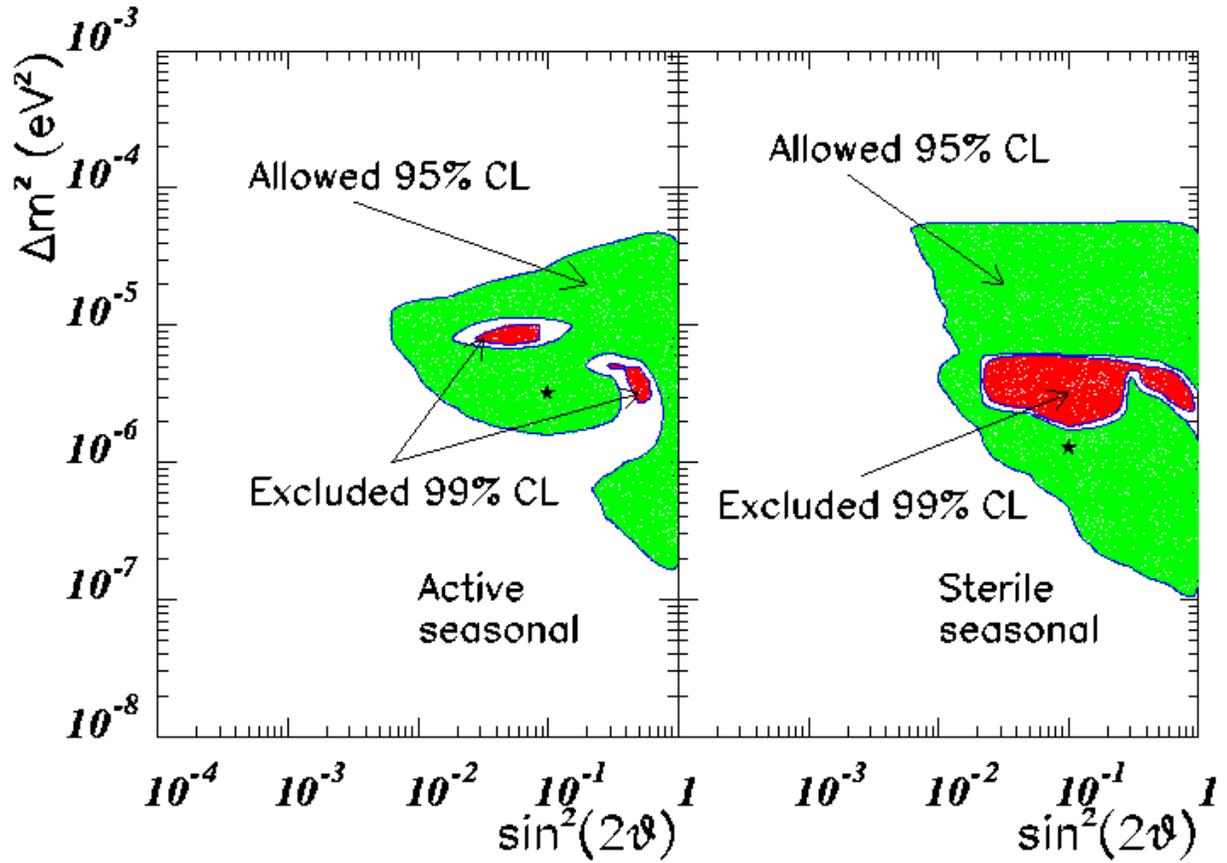,height=13cm}}
\end{center}
\vglue -.5cm
\caption{95 and 99\% CL regions obtained from the analysis of the
seasonal dependence of the event rates observed in Super--Kamiokande
from data with $E_e> 11.5 $ for active-active oscillations {\bf(a)}
and active-sterile oscillations {\bf(b)}. The larger area
represents the allowed region at 95\% CL. The darker area is the
excluded region at 99\% CL.}
\label{fig:season}
\end{figure}

In Fig.~\ref{fig:season} we show the exclusion regions at 95\% and
99\% CL. The larger area represents the {\sl allowed} region
at 95\% CL. Being an allowed region, it means that at 95\% CL, some
effect is observed. The darker area shows the small excluded region at
99\% CL.

Since the seasonal variation of the event rates in the MSW region is
due to neutrino regeneration in the Earth one expects a correlation
with the day-night variation which arises from the same origin. This
correlation is observed as the 95\% allowed regions in
Fig.~\ref{fig:season} have a large overlap with the 99\% excluded
region from the observed zenith angle dependence in
Fig.~\ref{fig:ratesdn}. Notice however that the upper part of the 
95\% CL allowed region for the oscillation in active flavours (larger
$\Delta m^2$ and larger mixing angles) is still in agreement with the
observed zenith angle dependence.  Thus, should the seasonal effect be
confirmed in the higher energy part of the spectrum, it will favour
the LMA solution to the solar neutrino problem.  

\subsection{Combined Analysis}
\label{fit:total}

We now present our results for the simultaneous fits to all the
available data and observables. In the combination we define the
global $\chi^2$ as the sum of the different $\chi^2$ defined above.
In principle such analysis should be taken with a grain of salt as
these pieces of information are not fully independent; in fact, they
are just different projections of the double differential spectrum of
events as a function of time and energy. Thus in our combination we
are neglecting possible correlations between the uncertainties in the
energy and time dependence of the event rates.

From the full data sample we obtain that the for the SSM of
Ref.~\cite{BP98}
\begin{equation}
\chi^2_{min}(SSM)=96.
\end{equation}
for 32 d.~o.~f. So the probability of explaining the full data sample
as a statistical fluctuation of the the SSM together with the SM of
particle interactions is smaller than $10^{-7}$.  

In the MSW oscillation region we obtain that for oscillations into
active flavours the best fits are obtained for the LMA solution with
\begin{eqnarray}
\Delta m^2 & = & 3.65\times 10^{-5} \mbox{eV}^2\; , \nonumber \\ 
\sin^2 2\theta & = & 0.79\;, \label{lmaglo} \\
\chi^2_{min}& = & 35.4 \; , \nonumber 
\end{eqnarray}
for 30 d.~o.~f. which implies that the solution is acceptable at 77\%
CL., while in the SMA region the local best--fit point is
\begin{eqnarray}
\Delta m^2 & =& 5.1\times 10^{-6} \mbox{eV}^2\; , \nonumber \\ 
\sin^2 2\theta& = & 5.5\times 10^{-3}\; , \label{smaglo} \\
\chi^2_{min}& = & 37.4  \;,\nonumber
\end{eqnarray}
valid at the 83\% CL.. The global analysis still presents a minimum 
in the LOW region with best--fit point:
\begin{eqnarray}
\Delta m^2 & = & 1.0\times 10^{-7} \mbox{eV}^2\; , \nonumber \\ 
\sin^2 2\theta & = & 0.94\;,  \label{lowglo} \\
\chi^2_{min}& = & 40. \; , \nonumber 
\end{eqnarray}
which is acceptable at 90\% CL. 

The results of the global analysis for the case of 
of active-sterile neutrino oscillations, show the fit is slightly worse
in this case than in the active-active oscillation scenario. The best--fit
point is obtained for the SMA solution with
\begin{eqnarray}
\Delta m^2 & =&  5.0\times 10^{-6} \mbox{eV}^2\; ,\nonumber \\
\sin^2 2\theta& = & 3.0\times 10^{-3}\; ,\label{steglob}\\
\chi^2_{min}& = & 40.2  \;,\nonumber
\end{eqnarray}
acceptable at 90\% CL. 
 
\begin{figure}
\begin{center}
\mbox{\epsfig{file=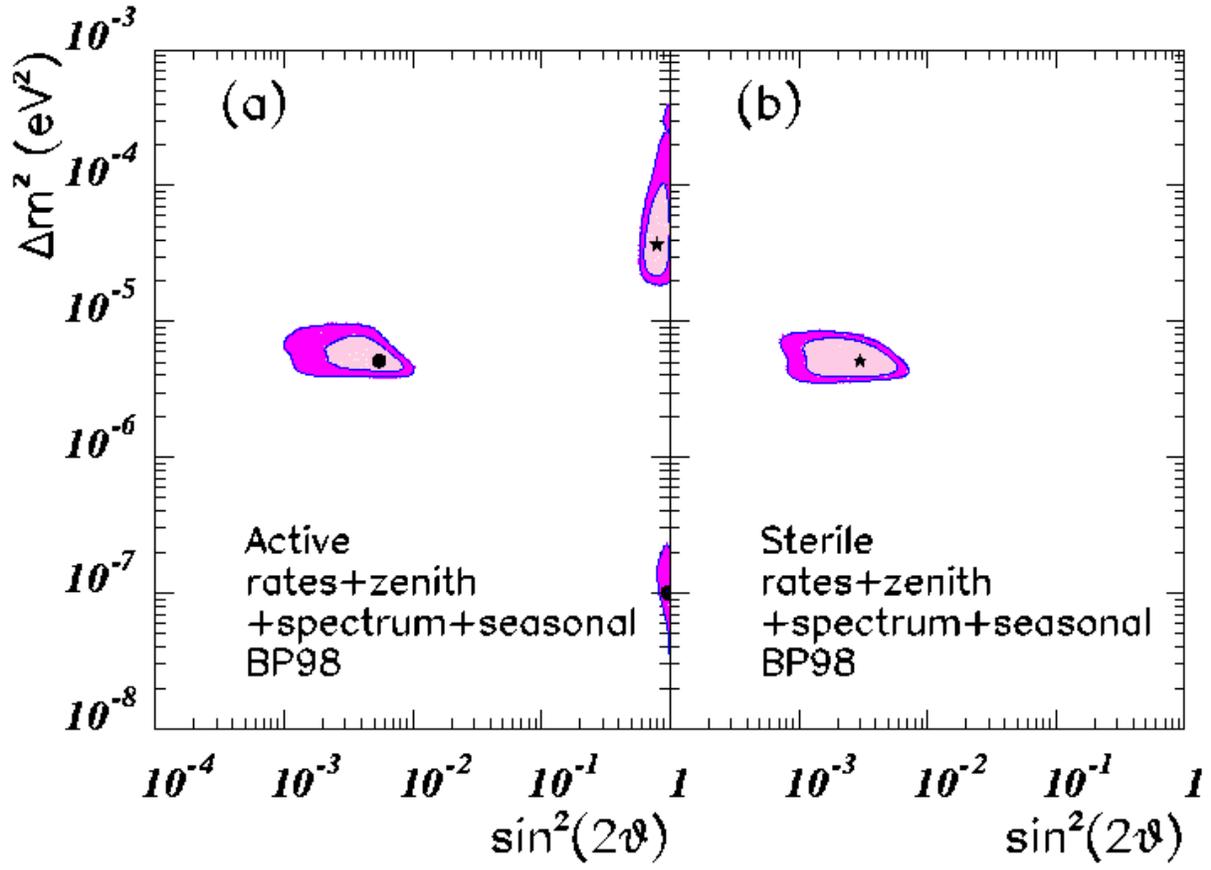,height=13cm}}
\end{center}
\vglue -.5cm
\caption{Allowed regions in $\Delta m^2$ and $\sin^2\theta$  from 
the measurements of the total event rates at the Chlorine, Gallium and
Super--Kamiokande (825-day data sample) combined with the zenith angle
distribution observed in Super--Kamiokande, the recoil energy spectrum
and the seasonal dependence of the event rates, for active-active
oscillations {\bf(a)} and active-sterile oscillations {\bf(b)} .  The
darker (lighter) areas indicate 99\% (90\%)CL regions.  
Global best--fit point
is indicated by a star. Local best--fit points are indicated by a dot.}
\label{fig:ratesdnss}
\end{figure}

In Fig.~\ref{fig:ratesdnss} we show the 90\% (lighter) and 99\%
(darker) CL allowed regions in the plane $\Delta m^2$-$\sin^2
2\theta$. Best--fit points in each regions are also indicated.

By comparing Figs.~\ref{fig:ratesdnss} and ~\ref{fig:rates} we see the
effect of the inclusion of the full data from Super-Kamiokande on both
time and energy dependence of the event rates.  For oscillations into
active neutrinos, the larger modification is in the LMA solution
region which has its lower part cut by the day-night variation data
while the upper part is suppressed by the data on the recoiled energy
spectrum. The best fit point is also shifted towards a larger mass
difference by a more than a factor 2.  As for the SMA solution region
the position of the best fit point is shifted towards a slightly
smaller mixing angle, while the size of the
region at the 90\% CL is reduced. At 99\% CL the allowed SMA region
is very little modified.  For oscillations into sterile neutrinos the
best fit point is also shifted towards a smaller mixing angle.

\begin{figure}
\begin{center}
\mbox{\epsfig{file=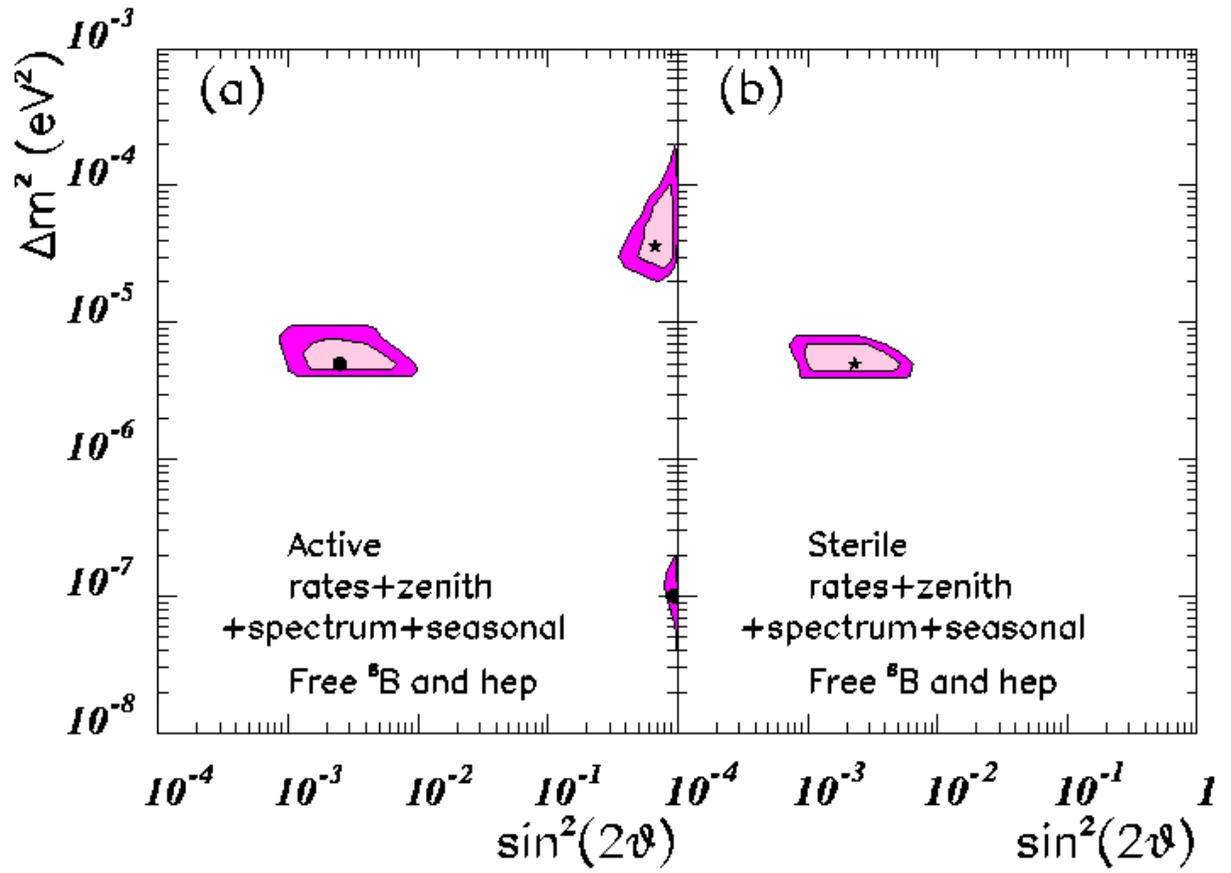,height=13cm}}
\end{center}
\vglue -.5cm
\caption{Same as previous figure allowing free $^8B$ and $hep$ 
neutrino flux normalizations.}
\label{fig:ratesdnssbo}
\end{figure}

We finally study the allowed parameter space with the normalization of
the $^8B$ and $hep$ fluxes are left free.  Figure
\ref{fig:ratesdnssbo} shows the allowed regions in the MSW parameter
space when the normalizations are allowed to take arbitrary values.
The best fit for the LMA solution occurs at
\begin{eqnarray}
\Delta m^2 & = & 3.6\times 10^{-5} \mbox{eV}^2\; , \nonumber \\
\sin^2 2\theta & = & 0.67\;,  \\
\beta=1.37& & \gamma=38  \nonumber \\ 
\chi^2_{min}& = & 30.7\; , \nonumber 
\end{eqnarray}
which is acceptable at 64\% CL. The best fit SMA solution is obtained
for
\begin{eqnarray}
\Delta m^2 & =&  5.0\times 10^{-6} \mbox{eV}^2\; ,\nonumber \\ 
\sin^2 2\theta& = & 2.5\times 10^{-3}\; , \\
\beta=0.61& & \gamma=12  \nonumber \\
\chi^2_{min}& = & 34  \;.\nonumber
\end{eqnarray}
which is acceptable at 80\% CL., The best fit for the LOW solution
occurs at
\begin{eqnarray}
\Delta m^2 & = & 1.0\times 10^{-7} \mbox{eV}^2\; , \nonumber \\
\sin^2 2\theta & = & 0.94\;,  \\
\beta=0.97& & \gamma=21  \nonumber \\ 
\chi^2_{min}& = & 38.1 \; , \nonumber 
\end{eqnarray}
which is acceptable at 90\% CL.,

In the case of active-sterile neutrino oscillations 
\begin{eqnarray}
\Delta m^2 & =&  5.0\times 10^{-6} \mbox{eV}^2\; ,\nonumber \\ 
\sin^2 2\theta& = & 2.0\times 10^{-3}\; , \\
\beta=0.61& & \gamma=12  \nonumber \\
\chi^2_{min}& = & 35.8  \;.\nonumber
\end{eqnarray}
which is acceptable at 85\% CL.,

For all solutions we find that the main effect of considering free 
normalization of the $^8B$ and $hep$ fluxes is upon the position of the 
best-fit points, mainly a reduction 
in the value of the mixing angle. The allowed regions are also
enlarged as can be seen by comparing Figs.~\ref{fig:ratesdnss} and 
\ref{fig:ratesdnssbo}.

\section{Summary and Discussion}
\label{conclu}

We have presented an updated global analysis of two-flavor MSW
solutions to the solar neutrino problem using the full data set
corresponding to the 825-day Super--Kamiokande sample plus Chlorine,
GALLEX and SAGE experiments. In addition to all measured total event
rates we included all Super--Kamiokande data on the zenith angle
dependence, energy spectrum and seasonal variation of the events.  We
have given a comparison of the quality of different solutions of the
solar neutrino anomaly in terms of MSW conversions of $\nu_e$ into
active and sterile neutrinos. For the case of conversions into active
neutrinos we have found that once the full data set is included both
SMA and LMA solutions give an equally good fit to the data.  We find
that the best--fit points for the combined analysis are $\Delta
m^2=3.6 \times 10^{-5}$ eV$^2$ and $\sin^2 2\theta=0.79$ with
$\chi^2_{min}=35.4/30$ d.~o.~f and $\Delta m^2=5.1~\times 10^{-6}$
eV$^2$ and $\sin^2 2\theta=5.5~\times 10^{-3}$ with
$\chi^2_{min}=37.4/30$ d.~o.~f.  In contrast with the earlier 504-day
study of Bahcall-Krastev-Smirnov our results indicate that the LMA
solution is slightly preferred. Although small, there is a hint for
seasonality in the data and this should therefore take into account in
future studies, as we have indicated here. Defining the seasonal
variation (in percent) as
\begin{displaymath}
Var \equiv 2 \frac{R^{th}_{max}-R^{th}_{min}}{R^{th}_{max}+R^{th}_{min}} 
\end{displaymath}
where $R^{th}_{max(min)}$ is the expected event rate during the winter
(summer) period, we find that the seasonal effect may still reach 11\%
in the lower part of the allowed LMA region (6\% comes from the
variation expected from the geometric effect due to the eccentricity
of the Earth's orbit), without conflict with the negative hints of a
day-night variation.  Such seasonal dependence is correlated with the
day-night effect and this in turn can be used in order to discriminate
the MSW from the vacuum oscillation solution to the solar neutrino
anomaly. We have performed a numerical study of this correlation
\cite{v99} which generalizes the estimate presented by
Smirnov at the 1999 edition of Moriond for a constant Earth density.
Our results show that due to the Earth matter profile the seasonal
variation can be substantially enhanced or suppressed as compared to
the expected value obtained with an average Earth density.  One must
notice, however, that the exact values of masses and mixing for which a
significant enhancement is possible may depend on the precise model of
the Earth density profile. The numerical results presented above were
obtained using the step function profile and thus numerical
differences with the results obtained with, for instance the PREM
model \cite{bks99} can be expected for a given point in the MSW plane.

With the results from the combined analysis one can also predict the
expected event rates at future experiments. For example, we find that
the average oscillation probabilities for $ ^7Be$ neutrinos in the
different allowed regions for MSW oscillations into active neutrinos
is
\begin{equation}
{P_{\rm ^7Be~SMA}} = 0.01^{+0.14}_{-0.004} ,
\end{equation}
\begin{equation}
{P_{\rm ^7Be~LMA}} = 0.55^{+0.08}_{-0.08} .
\end{equation}
These results imply that, at Borexino we expect a suppression on the
number of events as compared with the predictions of the SSM of
\begin{equation}
 R^{\rm SMA}/R^{\rm BP98}= 0.22^{+0.11}_{-0.004} ,
\end{equation}
\begin{equation}
 R^{\rm LMA}/R^{\rm BP98}= 0.65^{+0.06}_{-0.06} .
\end{equation}
For conversions into sterile neutrinos only the SMA solution is
possible with best--fit point $\Delta m^2=5.0~\times 10^{-6}$ eV$^2$
and $\sin^2 2 \theta=3.2 \times 10^{-3}$ and $\chi^2_{min}=40.2/30$
d.~o.~f which leads to
\begin{equation}
P_{\rm ^7Be~sterile} = 0.015^{+0.090}_{-0.002}, 
\end{equation}
and one expects a larger suppression of events at Borexino 
\begin{equation}
R^{\rm sterile}/R^{\rm BP98}= 0.015^{+0.090}_{-0.002}.
\end{equation}

As a way to improve the description of the observed recoil electron
energy spectrum we have also considered the effect of departing from
the SSM of Bahcall and Pinsonneault 1998 by allowing arbitrary $^8B$
and $hep$ fluxes.  Our results show that this additional freedom does
not lead to a significant modification of the oscillation parameters.
The best fit is obtained for $^8B/^8B_{SSM}=0.61$ and
$hep/hep_{SSM}=12$ for the SMA solution both for conversions into
active or sterile neutrinos and $^8B/^8B_{SSM}=1.37$ and
$hep/hep_{SSM}=38$ for the LMA solution.

\acknowledgments

M.~C. G.-G. is thankful to the Instituto de Fisica Teorica of UNESP
and to the CERN theory division for their kind hospitality during her
visits.  We thank comments by John Bahcall, Vernon Barger, Venya
Berezinsky, Plamen Krastev and Alexei Smirnov. This work was supported
by Spanish DGICYT under grant PB95-1077, by the European Union TMR
network ERBFMRXCT960090 and by Brazilian funding agencies CNPq, FAPESP
and the PRONEX program.

\begin{table}
\begin{tabular}{|l|l|l|l|l|}
Experiment & Rate & Ref. & Units& $ R^{\rm BP98}_i $\\
\hline
Homestake  & $2.56\pm 0.23 $ & \protect\cite{homestake} & SNU &  $7.8\pm 1.1 $   \\
GALLEX + SAGE  & $72.3\pm 5.6 $ & \protect\cite{gallex,sage} & SNU & $130\pm 7 $  \\
Super--Kamiokande & $2.45\pm 0.08$ & \protect\cite{sk700} & 
$10^{6}$~cm$^-2$~s$^{-1}$ & $5.2\pm 0.9 $ \\   
\end{tabular}
\vglue .3cm
\caption{Measured rates for the Chlorine, Gallium and Super--Kamiokande 
experiments. }
\label{rates}
\end{table}
\begin{table}
\begin{tabular}{|l|l|}
Angular Range & Data$_i\pm \sigma_{i}$ \\
\hline
Day $ 0<\cos\theta<1 $  & $0.463 \pm 0.0115$ \\
N1  $ -0.2<\cos\theta<0 $ & $0.512 \pm 0.026$\\
N2  $ -0.4<\cos\theta<-0.2 $ & $0.471 \pm 0.025$\\
N3  $ -0.6<\cos\theta<-0.4 $ & $0.506 \pm 0.021$\\
N4  $ -0.8<\cos\theta<-0.6 $ & $0.484 \pm 0.023$\\
N5  $ -1<\cos\theta<-0.8 $ & $0.478 \pm 0.023$
\label{daynight}
\end{tabular}
\vglue .3cm
\caption{Super--Kamiokande Collaboration zenith angle distribution 
of events \protect\cite{sk700}.}
\end{table}
\begin{table}
\begin{tabular}{|l|l|l|l|l|}
Energy bin & Data$_i \pm \sigma_{i,stat}$ & 
$\sigma_{i,exp}$ (\%) & $\sigma_{i,cal}$ (\%) & $\sigma_{i,uncorr}$ (\%) \\
\hline
    5.5 MeV $< E_e <6 $ MeV & $0.472 \pm 0.037$ 
    &  1.3 &  0.3  & 4.0   \\
    6 MeV $< E_e <6.5 $ MeV & $0.444 \pm 0.025$ 
    &  1.3 &  0.3  & 2.5  \\
    6.5 MeV $< E_e <7 $ MeV & $0.427 \pm 0.022$ 
    & 1.3  & 0.3  & 1.7  \\
    7 MeV $< E_e <7.5 $ MeV & $0.469 \pm 0.022$ 
    & 1.3  & 0.5  & 1.7  \\
    7.5 MeV $< E_e <8 $ MeV & $0.516 \pm 0.022$ 
    & 1.5  & 0.7  & 1.7  \\
    8 MeV $< E_e <8.5 $ MeV & $0.488 \pm 0.025$ 
    & 1.8  & 0.9  & 1.7  \\
    8.5 MeV $< E_e <9 $ MeV & $0.444 \pm 0.025$ 
    & 2.2  & 1.1   & 1.7 \\
    9 MeV $< E_e <9.5 $ MeV & $0.454 \pm 0.025$ 
    & 2.5  &  1.4 & 1.7  \\
    9.5 MeV $< E_e <10 $ MeV & $0.516 \pm 0.029$ 
    & 2.9   & 1.7  & 1.7  \\
    10 MeV $< E_e <10.5 $ MeV & $0.437 \pm 0.030$ 
    &3.3   & 2.0  & 1.7  \\
    10.5 MeV $< E_e <11 $ MeV & $0.439 \pm 0.032$ 
    &3.8   & 2.3  & 1.7   \\
    11 MeV $< E_e <11.5 $ MeV & $0.476 \pm 0.035$ 
    &4.3   & 2.6  & 1.7  \\
    11.5 MeV $< E_e <12 $ MeV & $0.481 \pm 0.039$ 
    &4.8   & 3.0  & 1.7   \\
    12. MeV $< E_e <12.5 $ MeV & $0.499 \pm 0.044$ 
    &5.3   & 3.4  & 1.7  \\
    12.5 MeV $< E_e <13 $ MeV & $0.538 \pm 0.054$ 
    & 6.0  & 3.8  & 1.7   \\
    13 MeV $< E_e <13.5 $ MeV & $0.530 \pm 0.069$ 
    & 6.6  & 4.3  & 1.7  \\
    13.5 MeV $< E_e <14 $ MeV & $0.689 \pm 0.092$ 
    &  7.3 & 4.7  & 1.7  \\
    14 MeV $< E_e<20  $ MeV  & $0.612 \pm 0.077$ 
    & 9.2  & 5.8  &  1.7  
\end{tabular}
\vglue .3cm
\caption{Recoil energy spectrum of solar neutrinos 
from the 825-day Super--Kamiokande Collaboration data sample
\protect\cite{sk700}. Here $\sigma_{i,stat}$ is the statistical error,
$\sigma_{i,exp}$ is the error due to correlated experimental errors,
$\sigma_{i,cal}$ is the error due to the calculation of the expected
spectrum, and $\sigma_{i,uncorr}$ is due to uncorrelated systematic
errors.}
\label{spectrum}
\end{table}
\begin{table}
\begin{tabular}{|l|l|}
Period (month) & Data$_i\pm \sigma_{i}$ $(E>11.5)$  \\
\hline
 $ 0.0<t<1.5 $  & $0.588 \pm 0.057$  \\
 $ 1.5<t<3.0 $  & $0.588 \pm 0.057$  \\
 $ 3.0<t<4.5 $  & $0.532 \pm 0.069$  \\
 $ 4.5<t<6.0 $  & $0.392 \pm 0.059$  \\
 $ 6.0<t<7.5 $  & $0.473 \pm 0.059$  \\
 $ 7.5<t<9.0 $  & $0.521 \pm 0.065$  \\
 $ 9.0<t<10.5 $  & $0.548 \pm 0.065$ \\
 $ 10.5<t<12.0 $  & $0.522 \pm 0.058$ 
\end{tabular}
\vglue .3cm
\caption{Seasonal distribution of events given by the Super--Kamiokande 
Collaboration  \protect\cite{sk700}.} 
\label{seasonal}
\end{table}
%
\begin{figure}
\begin{center}
\mbox{\epsfig{file=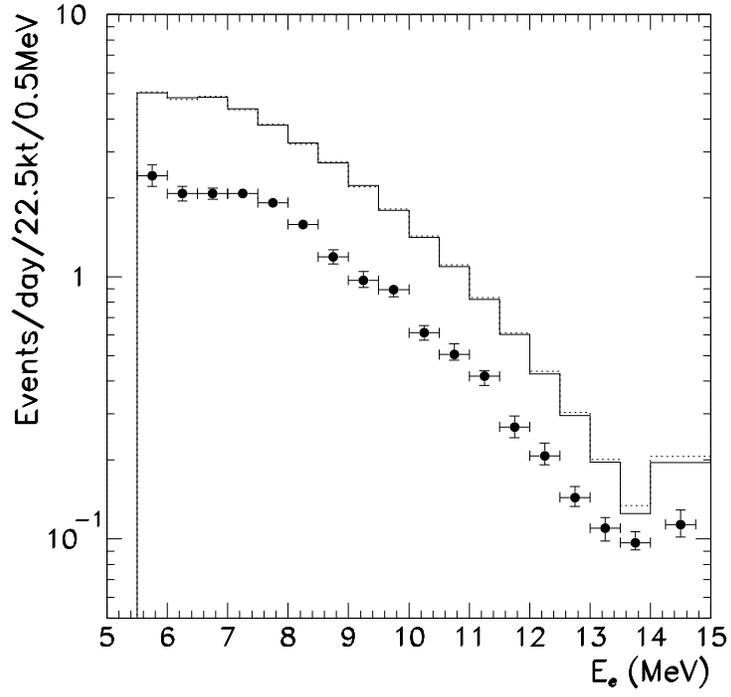,height=10.5cm}}
\end{center}
\caption{Theoretical recoil electron energy spectrum obtained by 
our calculation in Eqs. (\protect\ref{ratesth}), (\protect\ref{sigma}) (solid
histogram) compared with the Super--Kamiokande MC predictions (dotted
histogram).  Also shown are the data points from 
Super--Kamiokande data \protect\cite{sk700}.}
\label{fig:specmc}
\end{figure}
\end{document}